\documentclass[twoside,a4paper,12pt]{article}
\usepackage{amsmath}
\usepackage{amsfonts}
\usepackage{amssymb,amscd}
\usepackage{epsfig}
\usepackage{color}

\newcommand{\bea}{\begin{eqnarray}}
\newcommand{\eea}{\end{eqnarray}}
\newcommand{\bp}{\begin{pmatrix}}
\newcommand{\ep}{\end{pmatrix}}

\newcommand{\Slash}[1]{#1 \hspace{-.5em}/\hspace{.11em}}
\newcommand{\fourint}[1]{\int\!\frac{d^4 #1}{(2\pi)^4}}

\newcommand{\vect}[1]{{\mbox{\boldmath $#1$}}}

\newcommand{\dpart}[2]{\frac{\partial #1}{\partial #2}}

\begin{document}

\begin{flushright}
 ADP-02/69-T509 \\
 nucl-th/0203073
\end{flushright}
\medskip

\begin{center}
\begin{large}
{\bf Nucleon mass and pion loops: Renormalization
 $^\dagger$}\\
\end{large}
\vspace{1cm}
{\bf
M.~Oettel$^*$ and A.~W.~Thomas
\\}
\vspace{0.2cm}
Special Research Centre for the Subatomic Structure of Matter, 
University of Adelaide, Adelaide SA 5005, Australia 
\end{center}
\vspace{1cm}
\begin{abstract}
\normalsize
  Using Dyson--Schwinger equations, the nucleon propagator is analyzed
  nonperturbatively
  in a field--theoretical model for the pion--nucleon interaction. 
  Infinities are circumvented by using pion--nucleon form factors which define
  the physical scale. It is shown that the correct, finite,
  on--shell nucleon renormalization
  is important for the value of the mass--shift and the propagator.   
  For physically acceptable forms of the  pion--nucleon form factor
  the rainbow approximation together with renormalization is inconsistent. 
  Going beyond the rainbow approximation,
  the full pion--nucleon vertex is modelled by its bare part plus
  a one--loop  correction including an effective $\Delta$. 
  It is found that a consistent value for the nucleon mass--shift
  can be obtained as a consequence 
  of a subtle interplay between
  wave function and vertex renormalization.
  Furthermore, the bare
  and renormalized pion--nucleon coupling constant are approximately
  equal, consistent with results from the Cloudy Bag Model.

\end{abstract}

\vfill

\noindent
\rule{5cm}{.15mm}

\noindent
$^\dagger$Supported by a Feodor--Lynen fellowship of
the Alexander-von-Humboldt foundation and the Australian Research Council.\\
$^*$Address after April 30: MPI f\"ur Metallforschung, 
Heisenbergstr. 1, 70569 Stuttgart, Germany
\\

\newpage

\section{Introduction}
\label{int-sec}

The interest in a  consistent analysis of the effects of the pion cloud 
on nucleon properties is nurtured by two areas of recent research.
For one, the extrapolation of nucleon lattic data, quenched or unquenched,
to physical values of the quark (or pion) mass is mainly determined
by pionic effects 
\cite{Leinweber:1998ej,Young:2001nc}. Furthermore, these
effects will pose constraints on the development of covariant
nucleon models such as those presented in Refs.~\cite{Ishii:bu,Oettel:2000jj}.
The most basic observable of interest is the nucleon mass or, more precisely,
the nucleon mass shift associated with the pion cloud.
In a previous study \cite{Hecht:2002ej}, the connection of covariant Euclidean 
nucleon mass shift calculations to extant results from the Cloudy Bag Model
(CBM) \cite{Thomas:1981vc} and Chiral Perturbation Theory ($\chi$PT) 
\cite{Langacker:1973hh,Becher:1999he}
was analyzed. A nonperturbative analysis using the Dyson--Schwinger
(DS) equation in rainbow approximation (with no renormalization)
suggested that almost all of the nucleon mass shift can be attributed
to the covariant one--loop pion dressing. We will extend this analysis 
by supplementing the DS equation with correct on--shell renormalization
conditions (Section \ref{model-sec}). 
To properly implement those, one has to forgo 
an angular approximation employed in Ref.~\cite{Hecht:2002ej}. In
Section \ref{vert-sec} we calculate a one--loop correction 
to the $\pi NN$ vertex
which is subsequently employed in the DS equation. In the last section,
we summarize and present our conclusions.

\section{The model}
\label{model-sec}

We consider a pseudovector Lagrangian for the $\pi N$ interaction which 
reads in Euclidean space
\bea
{\cal L}_{\pi NN} &=& \frac{g}{2M_N} \;
   \bar \Psi\; i\gamma^\mu \gamma_5\;\vect \tau
  \Psi \cdot (\partial_\mu \vect \pi) \; .  
\eea
The DS equation for the nucleon propagator $G$ is given by
\bea
  \label{DSdef}
 G^{-1}(p) &=& G_0^{-1}(p) + i\Slash{p}\Sigma_V(p^2) + \Sigma_S(p^2) \; , \\
i\Slash{p}A(p^2)+ B(p^2) &=& Z_2\;(i\Slash{p} + Z_M M) + Z_1
   \fourint{k} \Gamma^0 \;G(k)\; \Gamma\; D(p-k) \nonumber \; , \\
  \Gamma^0 &=& (\Slash{p}-\Slash{k})\gamma_5\;\vect \tau\;g\;.
\eea
Because of the small mass of the pion, we can approximate the full
pion propagator $D(q)$ by the free scalar propagator,
$D^0(q)=(q^2+m_\pi^2)^{-1}$.
$\Gamma^0$ and $\Gamma$ stand for the free and the full 
$\pi NN$ vertex. The renormalization constants $Z_1$, $Z_2$ and $Z_M$
refer to the $\pi NN$ vertex, nucleon wave function and nucleon mass
respectively. The relation to bare quantities is given by
\bea
 g = \frac{Z_2}{Z_1}\;g_{\rm bare}\; , \qquad M_N = \frac{M_{\rm bare}}{Z_M} \; .
\eea
To account for the compositeness of the particles, we introduce a form 
factor at each $\pi NN$ vertex into this
idealized field--theoretic model, 
\bea
 \Gamma^{[0]}(p,k) &\to& \Gamma^{[0]}(p,k)\;u((p-k)^2) \;. 
\eea
In the case where both nucleons are on shell ($p^2=k^2=-M_N^2$), this form 
{}factor is most naturally related to the axial form factor 
of the nucleon. It is 
commonly parameterized as a dipole,
\bea
 \label{dip}
 u(q^2) &=& \left( \frac{\lambda^2-m_\pi^2}{\lambda^2+q^2}\right)^2 \;,
\eea
but we will also investigate (as in Ref.~\cite{Hecht:2002ej}) 
the exponential form
\bea
 u_{\rm exp}(q^2) &=& \exp\left( -\frac{q^2+m_\pi^2}{\Lambda^2} \right) \; .
\eea
From neutrino scattering experiments,  
the dipole width parameter (in the on--shell case) is 
determined as $\lambda=1.03 \pm 0.04$
GeV \cite{Thomas:kw}. This sets roughly the scale for our considerations.

Even after the introduction of form factors, we assume that it is sensible
to extract nonperturbative information from the DS equation 
(\ref{DSdef}). We note that now the loop integral in the equation is
convergent and all renormalization constants will therefore be finite.
A commonly used approximation to it is the rainbow approximation
$\Gamma = \Gamma^0$. The full $\pi NN$ vertex $\Gamma$ fulfills
its own DS equation which takes the symbolic form:
\bea
  \Gamma &=& Z_1\;\Gamma^0 + \Gamma^0\;(G^{[1]}G^{[2]})\; K_4\; \; .
\eea
Here, $K_4$ is the full, off-shell, amputated $N-N$ scattering matrix.
One sees that the rainbow approximation neglects the last term and enforces
$Z_1=1$.

We employ on--shell renormalization for the unknown nucleon propagator,
$G$. This requires that $G$ has a pole at $p^2=-M_N^2$ (the physical mass), 
\bea
 M_N(Z_2+\Sigma_V(-M_N^2)) &=& \Sigma_S(-M_N^2)+ Z_2 Z_M M_N \;, \\
 M_{\rm bare}-M_N = M_N(Z_M-1) &=& \frac{1}{Z_2}(M_N\Sigma_V(-M_N^2) - \Sigma_S(-M_N^2))
   \; . \label{dM}
\eea
The renormalization constant $Z_2$ is determined by the condition
that the residue at the pole be unity,
\bea
 \left. \dpart{G^{-1}}{(\Slash{ip})}\right|_{\Slash{ip}=-M_N} &=& 1 \quad \to \\
 -\Sigma_V(-M_N^2) +2M_N^2 \left.\dpart{\Sigma_V}{p^2}\right|_{p^2=-M_N^2}
  - 2M_N \left.\dpart{\Sigma_S}{p^2}\right|_{p^2=-M_N^2} &=& Z_2-1 \; . \quad
   \label{z2}
\eea
Mass renormalization ensures that the self--energy has cuts starting from the
physical thresholds. Furthermore one can show that the spectral
densities of the solution $G$ multiplied by $Z_2$ are properly normalized
(i.e. $Z_2$ can be interpreted as the probability of finding
a ``bare'' nucleon inside the pion--dressed one).

In the exploratory study \cite{Hecht:2002ej} the DS equation, Eq.~(\ref{DSdef}),
was solved in rainbow approximation 
after putting $Z_1=Z_2=Z_M=1$ and disregarding the above 
renormalization conditions. The mass of the dressed nucleon was then
found as the solution of $-M_D^2\;A(-M_D^2)+B(-M_D^2)=0$, $M_D<M_N$. 
In this simplified scenario it was possible to employ a certain angular 
approximation to the loop integral in eq.~(\ref{DSdef}) to calculate
all angular integrals analytically. However, if
one incorporates the renormalization
conditions (\ref{dM}) and (\ref{z2}), this angular approximation fails 
because it
underestimates the slopes of $\Sigma_V$ and $\Sigma_S$ considerably
(and these enter the expression for $Z_2$). As a result, 
we must resort to a numerical
computation of one angular integral. Since we have to 
evaluate the renormalization
conditions on the nucleon mass shell, it is necessary to continue
the DS equation to complex momenta. This intricate procedure is outlined 
in Appendix \ref{num-sec}.

\begin{figure}
 \begin{center}
  \epsfig{file=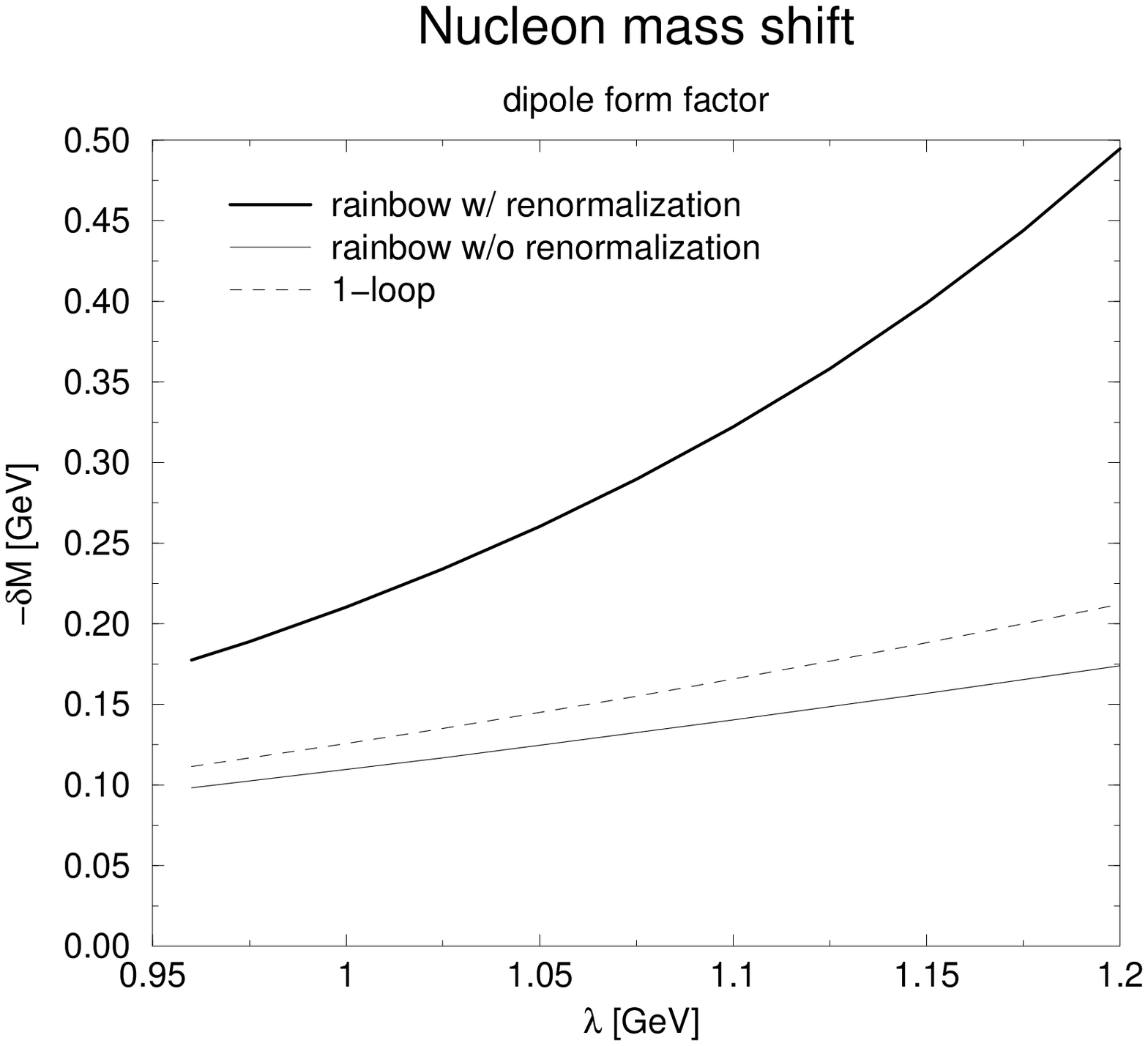,width = 6cm}
  \epsfig{file=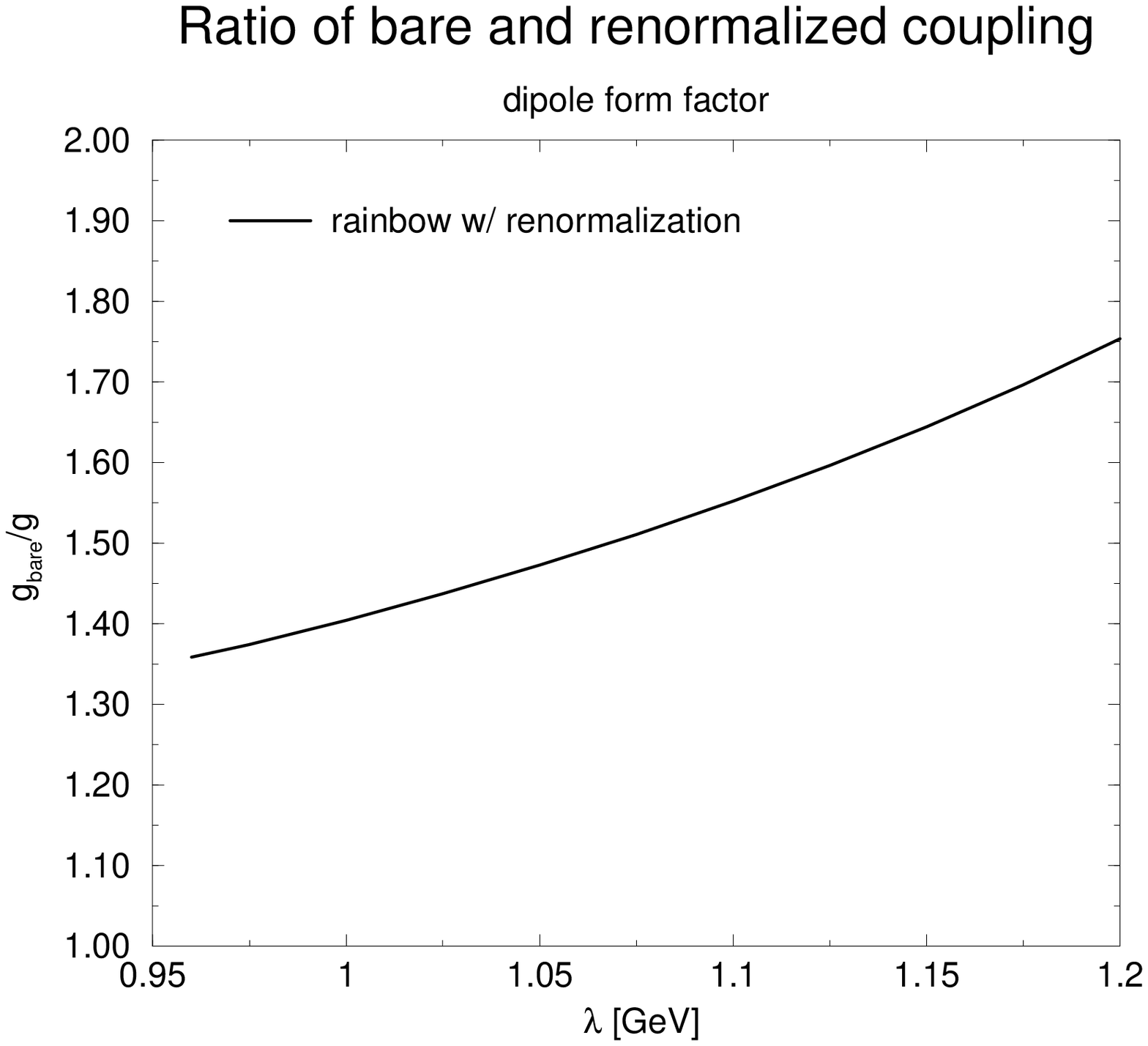,width = 6cm}
  \caption{Left panel: nucleon mass shift  in the renormalized treatment
  (defined as $\delta M = M_N-M_{\rm bare}$) and in the
   unrenormalized treatment (defined as $\delta M =M_D-M_N$) of the 
  rainbow approximation. 
  Right panel: the ratio
  $g_{\rm bare}/g = Z_1/Z_2$. }
  \label{rainbow-fig}
 \end{center}
\end{figure}

We illustrate the results for the mass shift in Fig.~\ref{rainbow-fig}, 
employing (as in Ref. \cite{Hecht:2002ej}) 
as the renormalized coupling constant $g=M_N/f_\pi$, i.e. $g_A=1$. We chose
a dipole form factor with cut-offs $\lambda$ in a range compatible 
with the measured axial form factor.\footnote{For technical reasons
we consider only
$\lambda>M_N$. Otherwise poles from the form factor enter the integration
domain, adding technical complications which are of no interest here.} 
Indeed, the mass shift for
the unrenormalized rainbow treatment and the one in one--loop approximation
are very close. In contrast, the mass shift in the renormalized rainbow
treatment is larger by a factor which rises from 1.8 ($\lambda =0.96$ GeV)
to 2.9  ($\lambda =1.2$ GeV). The explanation for this peculiar behaviour
can be found in the right panel of Fig.~\ref{rainbow-fig} where the
ratio of bare to renormalized coupling is plotted. This ratio equals
$1/Z_2$ in the rainbow treatment and is considerably larger than 1. If one
divides the DS equation, Eq.~(\ref{DSdef}),
by $Z_2$, one sees that the loop integral
is proportional to the bare coupling and therefore drives the mass shift
to larger absolute values.

A repetition of the calculations with the physical $g_A=1.26$ only 
aggravates the difference between renormalized and unrenormalized results.
Whereas the mass shift in the unrenormalized rainbow treatment
scales with $g_A^2$ (as the one-loop result), the mass shift in the 
renormalized case shoots up to values between 400 MeV and 1.9 GeV
(depending on $\lambda$), because of 
the nonlinear nature of the DS equation.

The question arises whether such a strong renormalization of the
pion--nucleon coupling constant, as visible in the rainbow solutions, 
is physically reasonable. An analysis
of pionic corrections to nucleons in the CBM to one--loop order
\cite{Thomas:1981vc,Theberge:1981mq} reveals that $Z_1 \approx Z_2$
(and therefore $g_{\rm bare} \approx g$) if one includes the $\Delta$ 
resonance in the analysis of the self--energies and the pion--nucleon
vertex.\footnote{The analysis there corresponds to a non--relativistic
expansion of positive--energy graphs in time--ordered perturbation 
theory.}
This is a strong indication that the rainbow approximation is unreliable
{}for this problem:
there $Z_1=1$ is set artificially and $Z_2 \sim 0.6 \dots 0.7$ for 
physical $\pi NN$ form factors.

Therefore, we will model the full $\pi NN$ vertex $\Gamma$ by the bare one 
plus a one-loop correction that includes the $\Delta$ in an effective manner,
i.e. as a spin--3/2 particle described by Rarita--Schwinger spinors.
Ideally, we would like to employ self--consistent propagators
of $N$ and $\Delta$ in this study, but solving a coupled system of two--loop
DS equations for $N$ and $\Delta$ is beyond the scope of this study.
Instead, we pursue a modest but nevertheless resource--consuming 
modification: the one--loop vertex correction
to $\Gamma$ will be calculated using free $N$ and $\Delta$ propagators
and the result will be inserted back into
the DS eq.~(\ref{DSdef}). By that token, the problem becomes numerically 
tractable, and we expect the bulk of  the effect on the solutions to be buried
in a reasonable estimate for $Z_1$. 

\section{Corrections to the $\pi NN$ vertex}
\label{vert-sec}

\begin{figure}[h]
  \epsfig{file=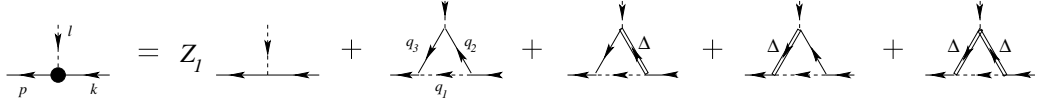,width=\textwidth}
  \caption{One--loop corrections to the $\pi NN$ vertex.}
  \label{v1loop-fig}
\end{figure}

We calculate the covariant one--loop correction
to the $\pi NN$ vertex by including the $\Delta$ as an effective degree
of freedom. Pictorially the equation for the vertex is displayed
in Fig.~\ref{v1loop-fig},
and as mentioned above we take for the nucleon and $\Delta$ propagators
free spin--1/2 and spin--3/2 propagators, respectively,
\bea
 G^0(p^2) &=& \frac{-i\Slash{p}+M_N}{p^2+M_N^2} \;, \\
 G_\Delta^{0,\mu\nu}(p^2) &=& \frac{-i\Slash{p}+M_\Delta}{p^2+M_\Delta^2}
 \;  {\mathbb P}^{\mu\nu}\;, \\
   {\mathbb P}^{\mu\nu} &=& \delta^{\mu\nu}
-\frac{1}{3}\gamma^\mu\gamma^\nu+
\frac{2}{3} \frac{p^\mu p^\nu}{M_\Delta^2} -
\frac{i}{3} \frac{p^\mu\gamma^\nu-p^\nu\gamma^\mu}{M_\Delta}\;,
\eea
with $M_N=0.94$ GeV and $M_\Delta=1.23$ GeV are the physical masses
of both particles. For the $\pi N\Delta$ and $\pi\Delta\Delta$ interactions
we employ tree level vertices derived from the simplest covariant
interaction Lagrangians, i.e.,
\bea
 \label{Lpnd}
 {\cal L}_{\pi N\Delta} &=& \frac{g_{\pi N\Delta}}{2M_N} \;
   \bar \Psi^\mu\; \vect T_{1/2}^{3/2}
  \Psi \cdot (\partial_\mu \vect \pi)  + {\rm h.c.} \to
   \Gamma^0_{\pi N\Delta} =\frac{g_{\pi N\Delta}}{2M_N}\;
  iq^\mu\;\vect T_{1/2}^{3/2} \; , \quad \\
 \label{Lpdd}
 {\cal L}_{\pi \Delta\Delta} &=& \frac{g_{\pi \Delta\Delta}}{2M_N}
  \bar \Psi^\rho\;i\gamma^\mu \gamma_5\;
 \vect T_{3/2}^{3/2}\;\Psi^\rho \cdot  (\partial_\mu \vect \pi)
 \to  \Gamma^0_{\pi \Delta\Delta} = \frac{g_{\pi \Delta\Delta}}{2M_N}\;
  \Slash{q}\gamma_5
 \;\vect T_{3/2}^{3/2} \; . \quad
\eea
Here $q$ is the momentum of the incoming pion. For the isospin 1/2--3/2
transition matrix we adopt the convention
\bea
  \vect T_{1/2}^{3/2} &=& C^{\frac{3}{2}M}_{\frac{1}{2}m,1k}\; .
\eea
As indices in the Clebsch--Gordan coefficient, $M,m,k$ stand for the isospin--$z$
components of $\Delta$, $N$  and $\vect \pi$. Furthermore, the isospin matrices
$\vect T_{3/2}^{3/2}$ are just the ones for the 4--dimensional $SU(2)$ representation.

For the numerical values of  the coupling constants $g_{\pi N\Delta}$
and $g_{\pi \Delta\Delta}$ we relate them to $g$ via $SU(6)$ quark model
expressions. This of course leaves room to fine--tuning which is not the main 
interest here. To apply the quark model, we note the non--relativistic
limit of both vertices,
\bea
 \Gamma^0_{\pi N\Delta} &\to& i\frac{g_{\pi N\Delta}}{2M_N}\;
  \vect T_{1/2}^{3/2} \; (\vect S_{1/2}^{3/2}\cdot \vect q)\;, \\\
  \Gamma^0_{\pi \Delta\Delta} &\to& -\frac{2i}{3} 
\;\frac{g_{\pi \Delta\Delta}}{2M_N}\;
  \vect T_{3/2}^{3/2} \; (\vect S_{3/2}^{3/2}\cdot \vect q)\; .
\eea
The transition matrix $\vect S_{1/2}^{3/2}$ is identical to
$\vect T_{1/2}^{3/2}$, it just refers to the spin degree of freedom.
Comparing to the expressions in Ref.~\cite{Thomas:1981vc}, we
readily find 
\bea
 g_{\pi N\Delta} &=& \sqrt{\frac{72}{25}}\;\;g\;, \\
 g_{\pi \Delta\Delta} &=& \frac{6}{5}\;g\;.
\eea

Here, we employ for the renormalized coupling constant the
physical value $g=g_A M_N/f_\pi$ with $g_A=1.26$.
Having fixed the strength of our interactions, we proceed now to the
calculation of the renormalization constant $Z_1$. We choose to fix it at the
virtual point where both nucleons and the pion are on shell, i.e.
\bea
 \label{z1fix}
 \bar u(k) \;\Gamma\; u(p) &\stackrel{!}{=}& \frac{g}{2M_N}\;
  \bar u(k) \;(\Slash{k}-\Slash{p})\gamma_5\; u(p)  \\
 && \left[p^2=k^2=-M_N^2,\; (p-k)^2=-m_\pi^2\right] \; .
\eea
This has the advantage that the $\pi NN$ form factor  
should reduce to 1 at this point.

\begin{figure}
 \begin{center}
  \epsfig{file=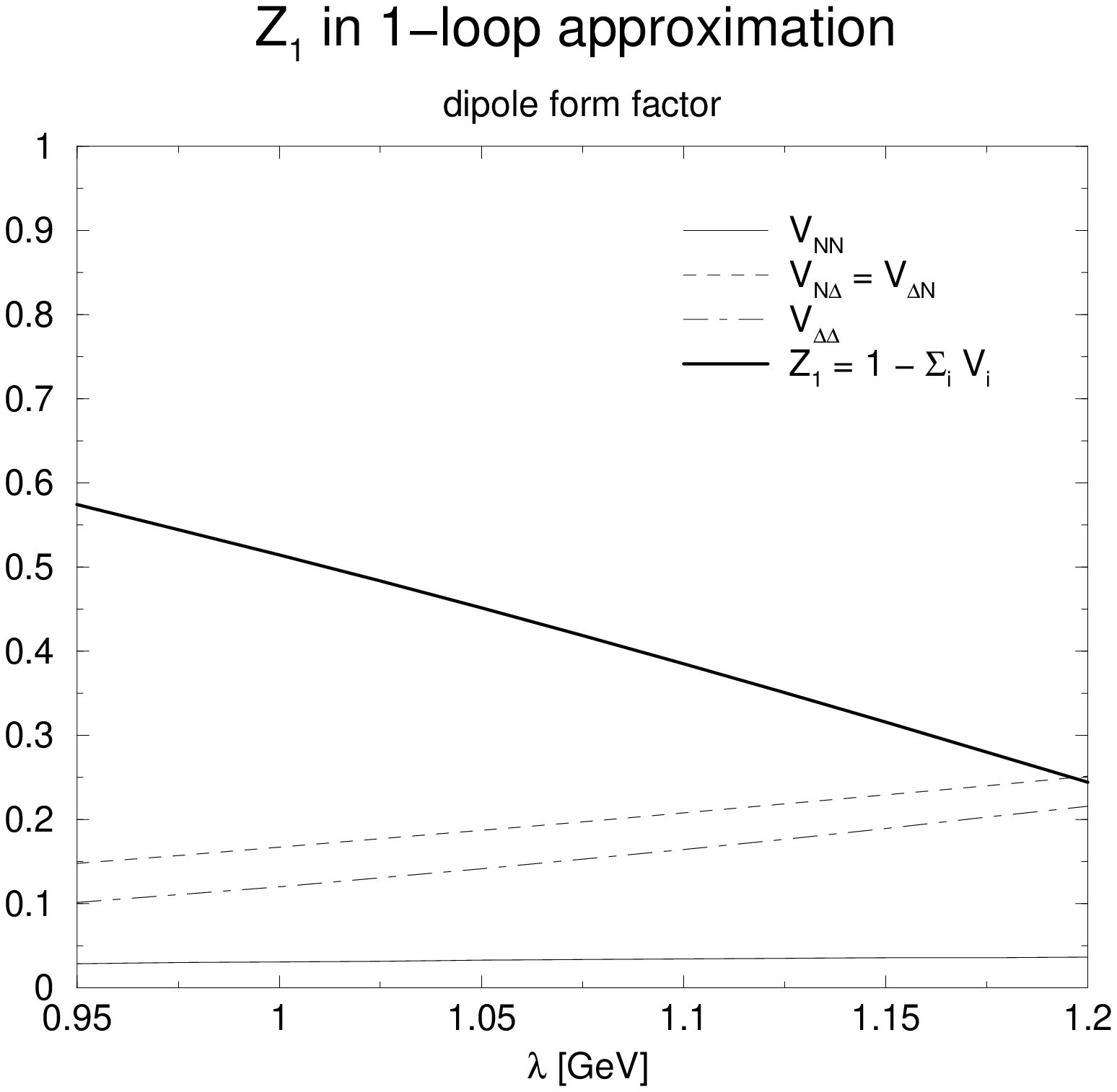,width = 6cm}
  \epsfig{file=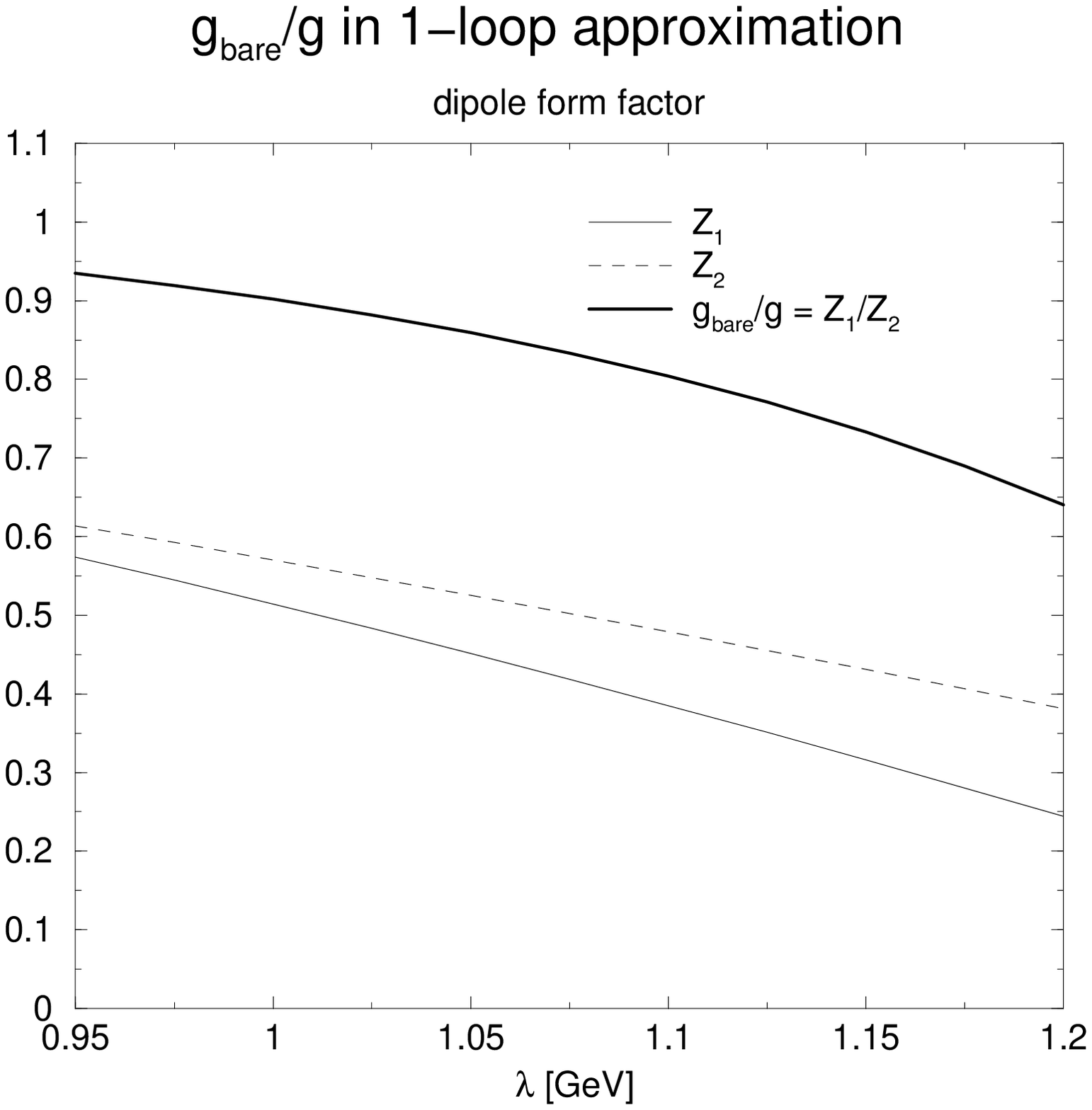,width = 6cm}
  \caption{Left panel: Vertex renormalization constant $Z_1$. 
  The different contributions to $1-Z_1$,
  stemming from the loop diagrams shown in
  Fig.~\ref{v1loop-fig}, are labelled with $V_{mn}$ with 
  $m,n \in\{N,\Delta\}$ labelling the baryons in the loop.
  Right panel: $Z_1$, $Z_2$ and their ratio in one--loop approximation.
  }
  \label{1loopv-fig}
 \end{center}
\end{figure}

The resulting $Z_1$ as a function of the dipole cut--off is plotted
in Fig.~\ref{1loopv-fig}. From the right panel one can see that
the loop--nucleons contribute less than 0.05 to $1-Z_1$. If this
were the whole story, the rainbow treatment of the DS equation would seem
to be justified, yielding mass-shifts $>500$ MeV as demonstrated before!
The bulk of the difference $1-Z_1$  comes from the two
graphs with one intermediate $\Delta$. The resulting bare coupling is
somewhat lower than the renormalized one. If the $\Delta$ were also included
in the nucleon self--energy, we would expect a somewhat larger bare
coupling since its contribution lowers $Z_2$ additionally. 
Keeping that in mind, we expect the present results for the
mass shift in the improved DS equation to be a lower bound.

Our one--loop improved model for the vertex can now be inserted into
the DS eq.~(\ref{DSdef}). Technically, we proceed by projecting the Dirac
structure of the vertex
onto basic covariant matrices,
\bea
 \Gamma_{\rm Dirac}(p,k,l) &=& (i\gamma_5)\;V_1(p^2,k^2,p\cdot k) +
    (\Slash{p}\gamma_5)\;V_2(p^2,k^2,p\cdot k) + \nonumber \\
   && (\Slash{k}_{\rm T}\gamma_5)\; V_3(p^2,k^2,p\cdot k) +
     (i\Slash{k}_{\rm T}\Slash{p}\gamma_5)\;V_4(p^2,k^2,p\cdot k) \; , \\
   k_{\rm T} &=& k - p\;\frac{k\cdot p}{p^2} \; ,
\eea
from which the scalar functions $V_{1\dots4}$ can be readily
traced out, using the program FORM \cite{Vermaseren:2000nd}.
The three remaining scalar integrals
are evaluated numerically. 
We calculate the functions $V_i$ on a three--dimensional
grid in the variables $p^2$, $k^2$ and $\hat p \cdot \hat k$, needed
for the solution of the DS equation.

\begin{figure}
 \begin{center}
  \epsfig{file=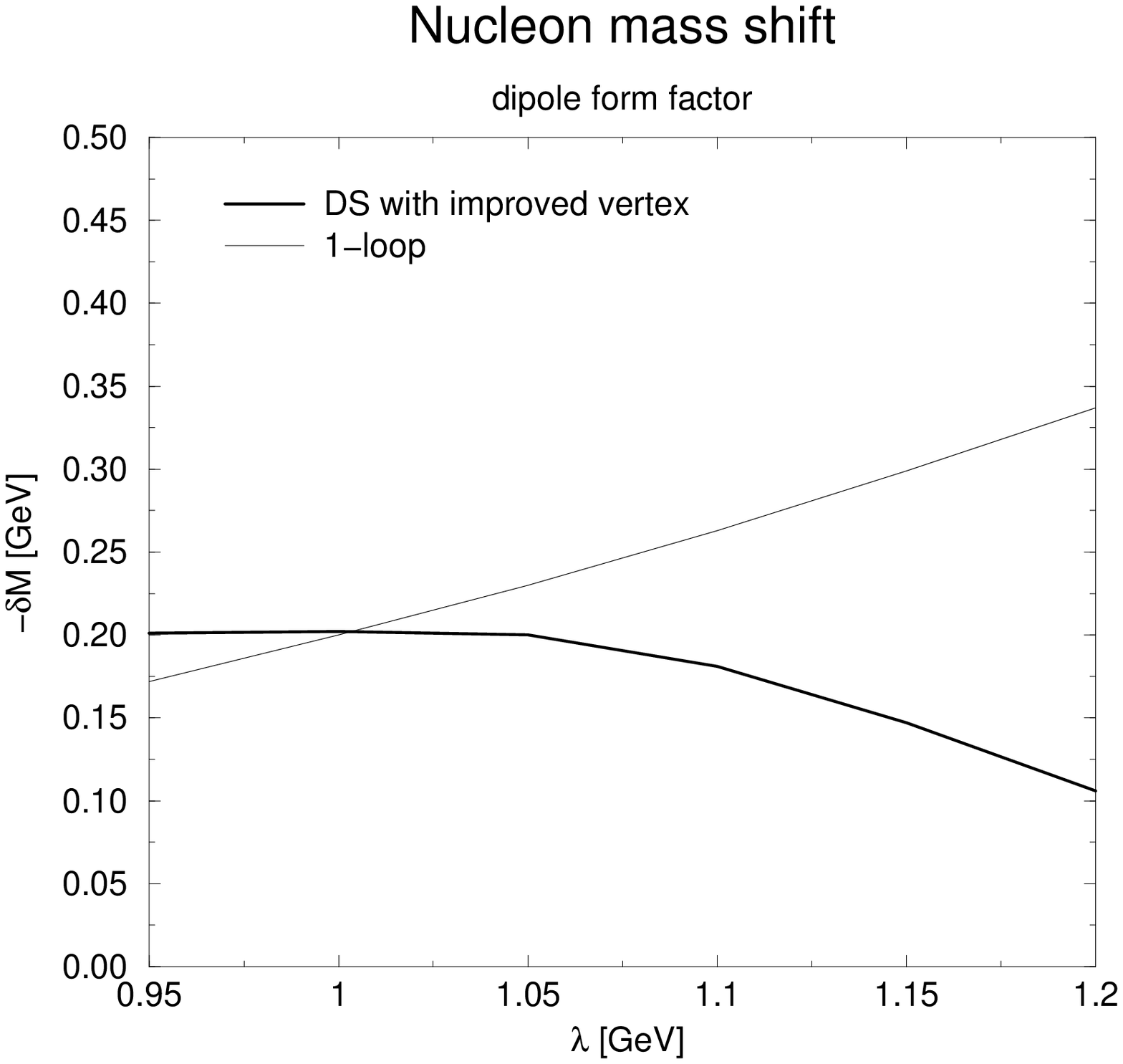,width = 6cm}
  \epsfig{file=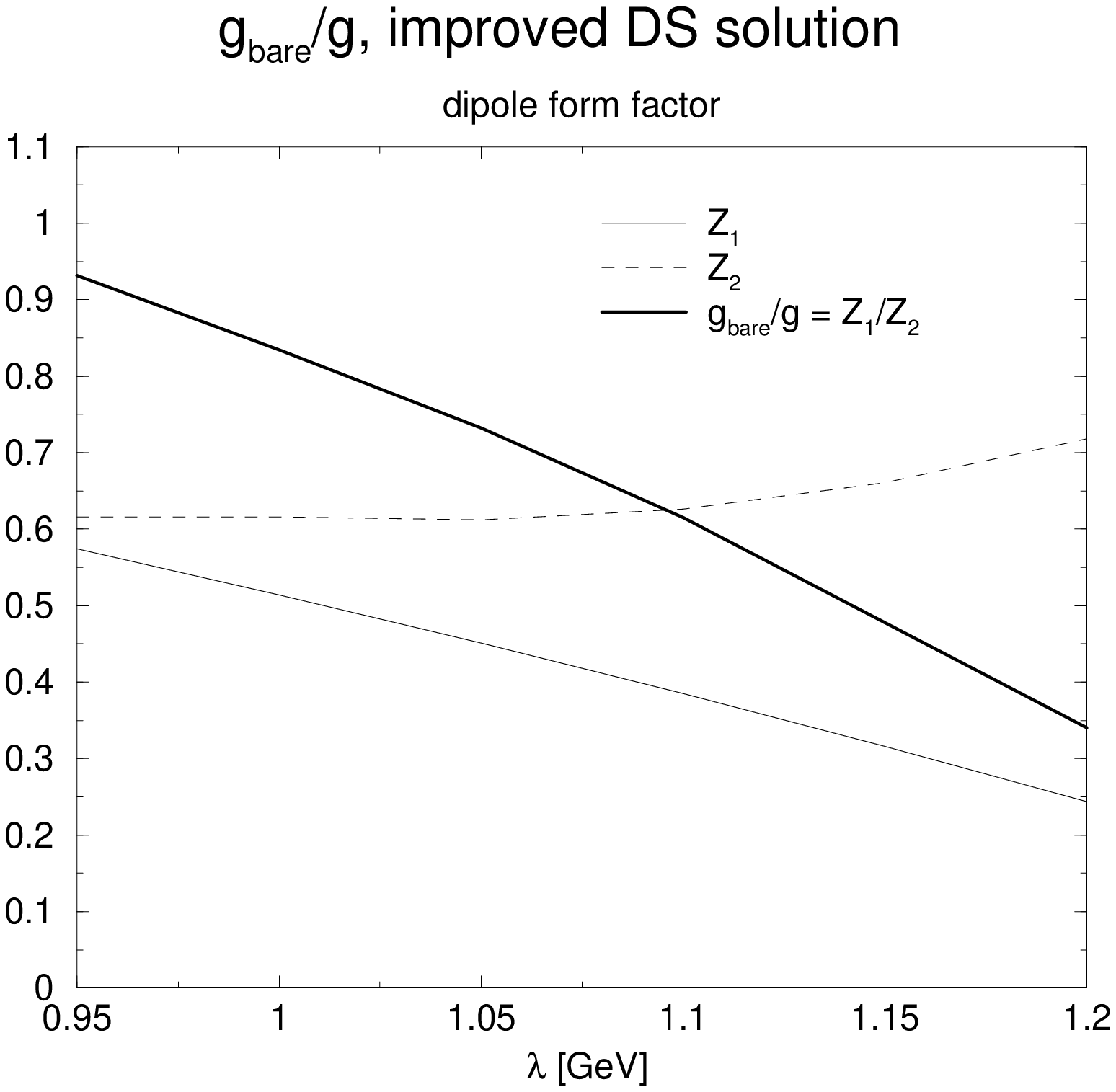,width = 6cm}
  \caption{Left panel: Mass shift for the vertex--improved DS solution
  in comparison to the one--loop approximation. Right panel: $Z_1$, $Z_2$ and 
  their ratio in the DS solution.
  }
  \label{improved-fig}
 \end{center}
\end{figure}

The results for mass-shift and the wave--function renormalization constant
$Z_2$ are depicted in Fig.~\ref{improved-fig}. In the physically interesting
region $\lambda \in [0.95,1.05]$ GeV the mass shift stays rather flat,
at a value around 200 MeV. For a harder form factor it actually
{\em drops}, which can again be explained by looking at the
ratio $g_{\rm bare}/g$: beyond $\lambda \sim 1.1$ GeV $g_{\rm bare}/g$ is 
becomes less than 1/2. As explained before, this ratio enters the pion loop
integral in the DS equation through $Z_1$. Since $Z_1$ itself is very low 
there, the one--loop treatment of the $\pi NN$ vertex can be questioned. 

The feature of a plateau in the mass shift for a certain range of cut--off
parameters (which happens to coincide with the experimental results
for the axial form factor) is actually a desired one in effective models
with cut--off functions, since it indicates the relative 
independence of the results on the
specific choice of cut--off. 
Whether that still holds after
self--consistent inclusion of the $\Delta$ resonance is an open question.
Certainly, by the arguments given above, its proper inclusion
should lead to $g_{\rm bare}/g>1$ and a larger mass shift.
Keeping this in mind, a value $-\delta M \approx 
-\delta M_{\rm 1-loop} \approx 200$ MeV constitutes a lower bound.

\begin{figure}
 \begin{center}
  \epsfig{file=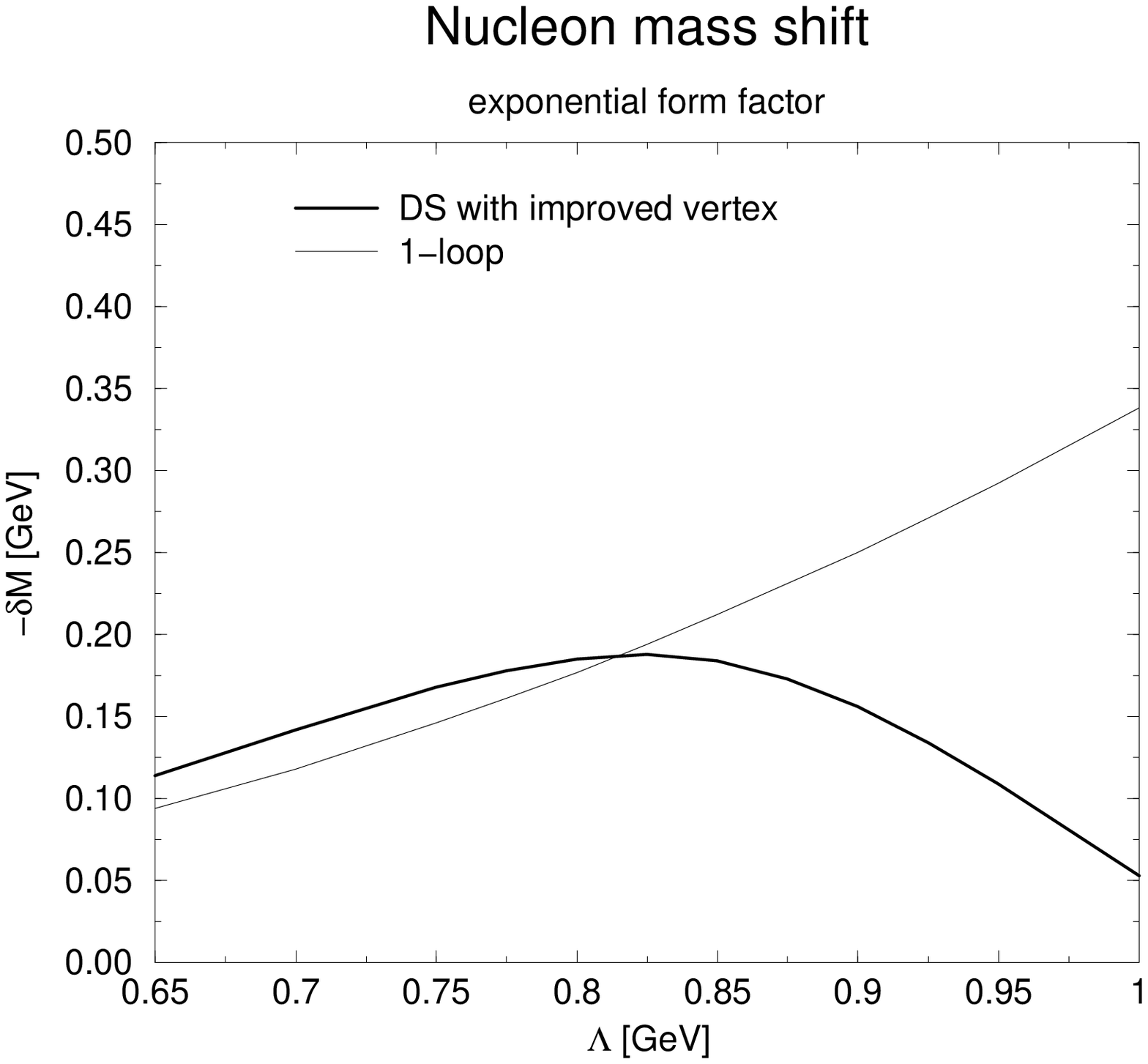,width = 6cm}
  \epsfig{file=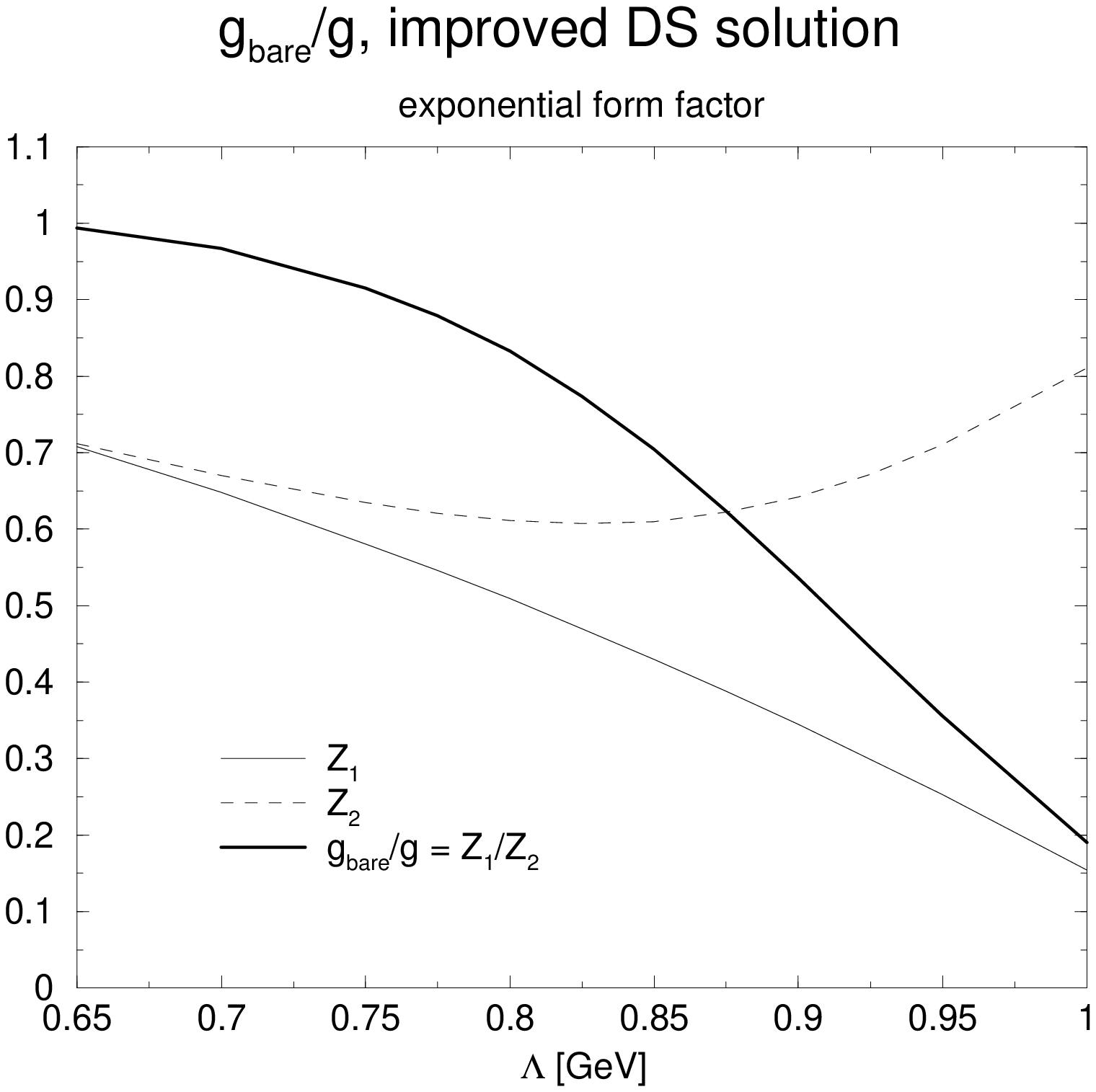,width = 6cm}
  \caption{Mass shift and renormalization constants arising from
 a pion--nucleon form factor of exponential type.
  }
  \label{improvedexp-fig}
 \end{center}
\end{figure}

In Fig.~\ref{improvedexp-fig} we show the results for an exponential
$\pi NN$ form factor. Since the exponential is an entire function,
we can investigate also somewhat softer form factors without encountering 
analytical problems. One--loop mass shifts are the same for dipole
and exponential, if $\lambda \approx 1.2 \Lambda$. The result for the 
self--consistent
mass--shift looks overall very similar to the previous case. Its curve  
flattens around $\Lambda = 0.8$ GeV at a value of 
approximately 190 MeV and then drops because of the rapidly
decreasing bare coupling constant.

\section{Summary and Conclusions}
\label{sum-sec}

We have investigated the covariant Dyson--Schwinger equation for 
the nucleon in a field--theoretic
model for the pion--nucleon interaction with a pseudovector La\-gran\-gian.
Its treatment in the commonly used rainbow approximation, including 
proper renormalization, leads to a strong renormalization of the
pion--nucleon coupling constant, $g_{\rm bare}/g >1$, 
in contradiction to perturbative one--loop
calculations. We were therefore led to calculate the one--loop perturbative
correction to the vertex, including the $\Delta$ resonance, and to re--solve
the DS equation. For physically reasonable values for the cut--off
in the dipole pion--nucleon form factor, $\lambda \approx 1$ GeV,
we find $g_{\rm bare}/g \lessapprox 1$ and a rather stable value for the
nucleon mass--shift of around $-200$ MeV, consistent with one--loop
results in both covariant treatments and semi--relativistic approaches 
such as the cloudy bag model. The fully self--consistent 
inclusion of the $\Delta$ resonance is expected to raise
the absolute value of the mass shift even further.

\section*{Acknowledgement}
This work was supported by the Australian Research Council, the
University of Adelaide and by the grant of a Feoder Lynen Fellowship
(M.O.) from the Alexander von Humboldt Foundation.

\begin{appendix}

\section{Solving the Dyson--Schwinger equation for complex momenta}
\label{num-sec}

Here we present a more detailed account of the procedure to
solve the DS eq.~(\ref{DSdef}),
\bea
  i\Slash{p}A(p^2)+ B(p^2) &=& Z_2\;(i\Slash{p} + Z_M M) + 
  i\Slash{p}\Sigma_V(p^2) + \Sigma_S(p^2) \;, \\
  i\Slash{p}\Sigma_V(p^2) + \Sigma_S(p^2) &=&  Z_1
   \fourint{k} \Gamma^0(p-k) \; \times \nonumber \\
    & & \quad \frac{1}{ i\Slash{k}A(k^2)+ B(k^2)}\; \Gamma(p,k) \; D^0(p-k) \; . \quad
\eea
In order to evaluate the renormalization conditions (\ref{dM},\ref{z2}),
we will need (as shown below) 
the functions $A(p^2)$ and $B(p^2)$ for $p^2\in[-(M_N-m_\pi)^2,\infty)$
-- i.e. if $p=(\vect 0,p_4)$, $p_4$ may assume  imaginary values.

\begin{figure}
 \begin{center}
   \epsfig{file=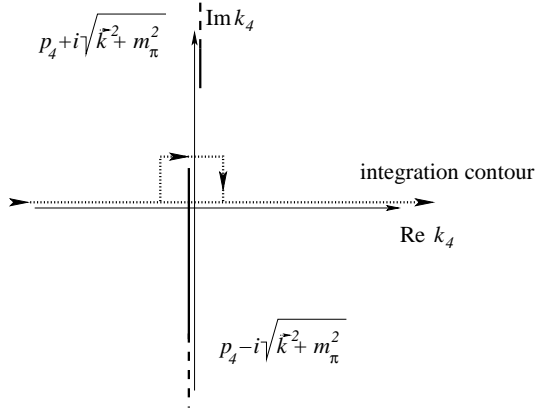,width=7cm}
   \caption{The $k_4$ integration in the pion loop. Once
  the singularities have crossed the real axis, the proper integration contour
  consist of the real axis and a loop contour around the singularities
  that have crossed the axis.}
  \label{res-fig}
 \end{center}
\end{figure} 

We introduce the two component vectors 
$\Sigma(p^2)=(\Sigma_S(p^2),\Sigma_V(p^2))$ and
$S(p^2)=(A(p^2),B(p^2))$. After applying suitable traces and doing the two 
trivial angular integrals we arrive at an equation of the form
\bea
 \Sigma_i (p^2) &=& \frac{3}{(2\pi)^3}\int_0^\infty \frac{k^2\;dk^2}{k^2A^2(k^2)+
 B^2(k^2)} \;\times \\
 & & \int_{-1}^1 \frac{\sqrt{1-z^2}\;dz}{p^2+k^2+m_\pi^2-2p_4|k|z}
  \; K_{ij}(p^2,k^2,z)\;S_j(k^2) \; , \nonumber
\eea
where $z=\cos{\psi}$ refers to the angle between the Euclidean vectors $p$ 
and $k$. If we want to evaluate the loop integral for $p^2<0$ 
($p_4$ imaginary), we have to note that for $p^2+m_\pi^2<0$ poles
from the pion propagator cross the real $k_4$ axis. These have to be avoided
by a suitable deformation of the integration contour. This situation is 
depicted in Fig.~\ref{res-fig}. Alternatively, the original real path 
may be retained provided that
a loop contour around the poles is added. This leads to
\bea
  \Sigma_i (p^2) &=& \frac{3}{(2\pi)^3}\int_0^\infty \frac{k^2\;dk^2}{k^2A^2(k^2)+
 B^2(k^2)} \;\times \nonumber\\
 & & \int_{-1}^1 \frac{\sqrt{1-z^2}\;dz}{p^2+k^2+m_\pi^2-2p_4k_4z}
  \; K_{ij}(p^2,k^2,z)\;S_j(k^2) \;  \nonumber +  \\
 &&  \frac{3}{(2\pi)^3}\; \theta(-p^2-m_\pi^2)\;\int_{\rm poles}\; 
  \frac{d^3\vect k}{2\sqrt{\vect k^2+m_\pi^2}}\; \times \nonumber\\
  &&\left. 
  \frac{K_{ij}(p^2,k^2,z)}{k^2A^2(k^2)+B^2(k^2)}\;S_j(k^2) 
  \right|_{(p-k)^2=-m_\pi^2}\; .
\eea
The loop contour in $k_4$ picks up 
residues $(-2\pi i)/(-2i\sqrt{\vect k^2+m_\pi^2})$ from the pion propagator,
and the integrand for the remaining integral in 3--space has to be evaluated
at the pion poles (thus the value of $z$ to be used in 
$K_{ij}$ is determined). We convert 
the integral over $\vect k$ into an integral over the squared four--momentum
$k^2$ using
\bea
  d\vect k^2 \frac{|\vect k|}{2\sqrt{\vect k^2+m_\pi^2}}& =& -dk^2 \;
   k^2 \;F(k^2,p^2) \;, \\
 F(k^2,p^2) &=& \frac{ \sqrt{(k^2+p^2+m_\pi^2)^2-4k^2p^2}}{4k^2p^2}\;.
\eea
Hence, we arrive at the final expression for the self--energies, valid also for
timelike momenta up to $-p^2=M_N^2$:
\bea
  \Sigma_i (p^2) &=& \frac{3}{(2\pi)^3}\int_0^\infty \frac{k^2\;dk^2}{k^2A^2(k^2
)+
 B^2(k^2)} \;\times \nonumber\\
 & & \int_{-1}^1 \frac{\sqrt{1-z^2}\;dz}{p^2+k^2+m_\pi^2-2p_4k_4z}
  \; K_{ij}(p^2,k^2,z)\;S_j(k^2) \;  \nonumber -  \\
  &&\frac{3}{(2\pi)^2}\; \theta(-p^2-m_\pi^2)\;
  \int_{-\left(\sqrt{-p^2}-m_\pi\right)^2}^{-p^2-m_\pi^2}
  \;k^2 F(k^2,p^2)\;dk^2\;\times \nonumber \\
  &&\left. 
  \frac{K_{ij}(p^2,k^2,z)}{k^2A^2(k^2)+B^2(k^2)}\;S_j(k^2) 
  \right|_{(p-k)^2=-m_\pi^2}\; .
\eea 
Now one sees clearly that in order to evaluate the renormalization
conditions at $p^2=-M_N^2$, one needs to know the self--consistent 
solution to $A(p^2)$ and $B(p^2)$ for the interval $p^2\in
[-(M_N-m_\pi)^2, \infty)$. The DS equation can now be 
solved iteratively, performing both $z$ and $k^2$ integrations numerically.
By virtue of the pion--nucleon form factors, the numerical treatment is not 
hampered by ultraviolet divergences and stable results are achieved
by using a mininum of 100 mesh points for each integral.

\end{appendix}

\end{document}